\documentclass[aps,prb,twocolumn,showpacs,superscriptaddress,preprintnumbers,amsmath,amssymb]{revtex4}
\usepackage{graphicx}
\usepackage{enumerate}
\usepackage{pstricks, tensors, multido}
\usepackage{nicefrac}
\usepackage{dcolumn}% Align table columns on decimal point
\usepackage{bm}% bold math

\psset{unit=1cm}
\definecolor{lightblue}{RGB}{30,144,255}
\definecolor{mediumgray}{gray}{0.65}
\definecolor{darkgreen}{rgb}{0,0.5,0}
\SpecialCoor% necessary for the postscript-style coordinates of form (! x y)
\def\plotwidth{8.66cm}
\newcommand{\cH}{\mathcal{H}}
\newcommand{\cHl}{\mathcal{H}_{\leftarrow}}%_{\mathfrak{L}}}
\newcommand{\cHr}{\mathcal{H}_{\rightarrow}}%_{\mathfrak{R}}}
\newcommand{\cHlr}{\mathcal{H}_{\leftrightarrow}}%_{\mathfrak{l,r}}}
\DeclareMathOperator{\crd}{crd}

\newcommand{\ie}{i.e.~}
\newcommand{\eg}{e.g.~}

\begin{document}

%\preprint{APS/123-QED}

\title{Unconstrained Tree Tensor Network:\\An adaptive gauge picture for enhanced performance}% Force line breaks with \\

\author{M. Gerster}
\affiliation{Institut f\"ur Quanteninformationsverarbeitung, Universit\"at Ulm, D-89069 Ulm, Germany}
\author{P. Silvi}
\affiliation{Institut f\"ur Quanteninformationsverarbeitung, Universit\"at Ulm, D-89069 Ulm, Germany}
\author{M. Rizzi}
\affiliation{Institut f\"ur Physik, Johannes-Gutenberg-Universitaet, D-55128 Mainz}
\author{R. Fazio}
\affiliation{NEST, Scuola Normale Superiore \& Istituto Nanoscienze CNR, I-56126 Pisa, Italy}
\affiliation{Centre for Quantum Technologies, National University of Singapore, 3 Science Drive 2, 117543 Singapore}
\author{T. Calarco}
\affiliation{Institut f\"ur Quanteninformationsverarbeitung, Universit\"at Ulm, D-89069 Ulm, Germany}
\author{S. Montangero}
\affiliation{Institut f\"ur Quanteninformationsverarbeitung, Universit\"at Ulm, D-89069 Ulm, Germany}

\date{\today}% It is always \today today, but any date may be explicitly specified

\begin{abstract}
We introduce a variational algorithm to simulate quantum many-body states
based on a tree tensor network ansatz which releases the 
isometry constraint usually imposed by the real-space renormalization coarse-graining:
This additional numerical freedom, combined with the loop-free topology of the tree network,
allows one to maximally exploit the internal gauge invariance of tensor networks, ultimately
leading to a computationally flexible and efficient algorithm 
able to treat open and periodic boundary conditions on the same footing.
We benchmark the novel approach against the 1D Ising model in transverse field with 
periodic boundary conditions and discuss the strategy to cope with the broken translational invariance 
generated by the network structure.
We then perform investigations on a state-of-the-art problem, namely the bilinear-biquadratic model in 
the transition between dimer and ferromagnetic phases. Our results clearly display an 
exponentially diverging correlation length and thus support the most recent guesses on the peculiarity of the transition.
\end{abstract}

% PACS, the Physics and Astronomy Classification Scheme.
%\keywords{Suggested keywords}%Use showkeys class option if keyword
% display desired
\maketitle

\section{Introduction}

Simulating quantum many-body states with tailored microscopical variational ans\"atze has been
refreshed in the last decade thanks to the introduction of tensor network states.
Despite being originally related~\cite{PrimoMPS1, PrimoMPS2, PrimoMPS3}
to density matrix renormalization group~\cite{UliMPS, GabriMPS, White92} schemes,
these variational architectures have been engineered to encompass a wide variety of physical
situations~\cite{Verstraetedenrimer,Completegraph},
thus widening the capabilities of the traditional numerical renormalization group (RG) approach.
Generally, tensor networks encode in a compact, numerically efficient way, many-body wavefunction
amplitudes over a real-space local basis expansion: the main reason for this real-space choice is
the fact that, since typical Hamiltonians are characterized by two-body interactions which decay 
sufficiently fast with the pairwise distance,
physically meaningful states (e.g.~ground states, lowest excited states, thermal states)
obey precise scaling laws on entanglement entropy under a real-space
bipartition~\cite{Arealaw1,Arealaw2,Arealaw3,Arealaw4}.
Such entanglement scaling can be precisely encoded in the tensor network
paradigm~\cite{BarthelScaling,BranchMeraScaling} and led to the design of 
various tensor network geometries, such as Matrix Product States (MPS)~\cite{PrimoMPS1},
PEPS~\cite{PowaPEPS}, Complete Graph states~\cite{CPScompletegraph}.

A physically sensible
class of tensor network architectures are the hierarchical (or holographic~\cite{JavierHolographic}) tensor networks:
they have the key feature of combining the usual local quantum space numerical renormalization together with a real-space
coarse graining, much like in the original RG picture by Wilson~\cite{Wilson75}.
Tree tensor networks (TTN)~\cite{VidalTTN,PsiTTN2010}, multiscale entanglement renormalization ansatz
(MERA) states~\cite{PrimoMERA,TTNevo2008,QuMERAchan2008}, and the recently-introduced branching MERA~\cite{BranchMERA},
are the most prominent examples of hierarchical tensor networks.
The fact that their network structure naturally embeds a scale invariance, makes them the ideal choice
for representing critical (gapless) quantum phases of matter, which are characterized by conformal invariance~\cite{Mussardobook}.
Moreover, hierarchical tensor networks can indeed satisfy the scaling rules of entanglement
of critical states~\cite{Cardyconformalentanglement}, both in those cases where area laws are logarithmically
violated (\eg in 1D critical systems) and in those where area laws are satisfied (\eg bosonic critical
systems in two or higher dimensions)~\cite{BranchMeraScaling}.

TTN states show a smooth computational
scaling with the tensor network bond dimension $m$ for the involved algebraic operations 
(\eg for binary TTN it is never higher than $\mathcal{O}(m^4)$).
This allows one to push numerical precision and description capabilities by sensitively increasing $m$,
making the TTN ansatz a potentially competitive method for simulating quantum many-body states.
On the other hand, TTN suffer more of a kind of entanglement clusterization, which is much more
alleviated in other approaches, such as MERA, thanks to the presence of disentangling operations
in their structure.

It is important to stress that the traditional scheme for simulating quantum lattice models with TTN
states~\cite{TagliacozTTN} relies on a particular selection of the internal 
tensor network gauge symmetry:  
in accordance to the RG flow picture by Wilson,
the tensors are fixed to be isometric operators in the real-space renormalization direction.
Although this gauge selection has indeed historical motivation, and moreover it guarantees some
useful mathematical properties in the thermodynamical limit
(namely, the complete positivity of the causal maps~\cite{QuMERAchan2008,PsiTTN2010, Evenbly2009, PrimoMERA}),
it confers no advantage in the simulation of finite-size systems, where actually it is more a hindrance.
Instead, if no ``a priori'' rigid selection of the isometric gauge from bottom to top is made,
one can always adjust the tensor network gauge to gain a computational enhancement in the algorithm.
This type of manipulation is particularly useful for tensor networks without closed loops in their
topology, like TTN, 
for which gauge flexibility translates into
a simplification of the variational algorithm into a simple eigenvalue problem 
(as it happens, for instance, for open boundary MPS compared to periodic ones).

In this manuscript we describe in detail an algorithm to find the ground state of quantum
lattice Hamiltonians, based on unconstrained (i.e.~gauge adaptive) tree tensor networks. 
We discuss thoroughly the computational cost scaling 
with numerical parameters, first of all the tensor network bond dimension $m$.  
We test the algorithm on one-dimensional quantum models, in both open (OBC) and periodic boundary conditions (PBC):
we use the quantum Ising model as a benchmark, and then investigate the bilinear-biquadratic spin-1 model
in proximity of the interface between the ferromagnetic and dimer phases.
In the latter, a peculiar exponential scaling of correlation lengths has been conjectured 
to explain the traditional toughness of the numerical problem. 
The nice agreement of our data with the most recent state-of-the-art calculations
corroborates the validity of the variational strategy presented here.
We also address the problem of restoration of translational invariance, which is broken by the TTN
architecture; we inquire for which physical quantities it is meaningful to average incoherently over translations
as opposed to local evaluation over a highly-entangled cluster of sites.
Ultimately, such investigation reveals quite a different behavior between local observations and
correlations.

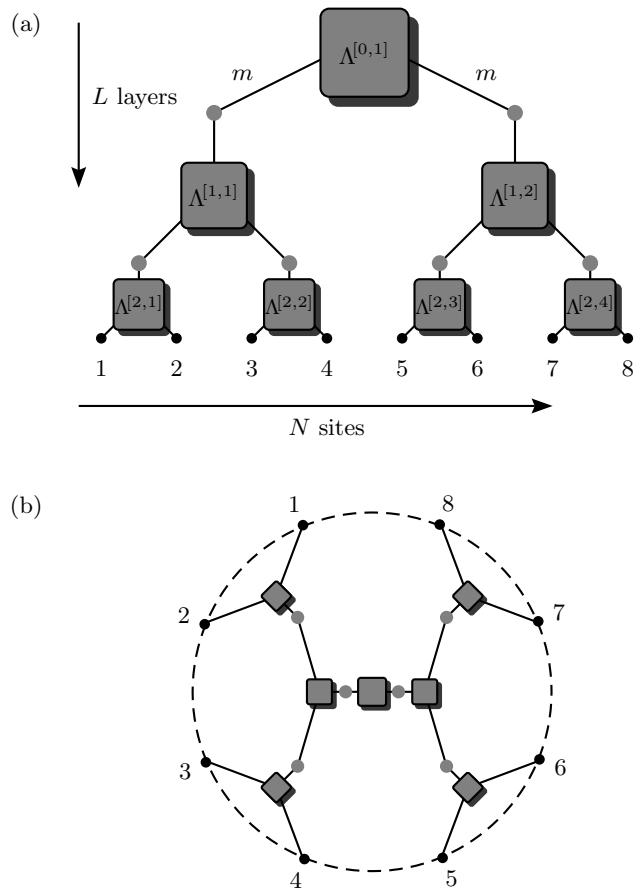
\begin{figure}
	\begin{pspicture}(-1.4,-1.5)(7.2,4.5)
	  \uput[-90](-1.0,4.5){ (a) }
	  
	  \multido{\nX=0.5+2.0, \iN=1+1}{4}{ 
	    \Tens[ \nX, 0.5, 0.5, 0.7, gray, \footnotesize{$\Lambda^{\![2,\iN]}$} ] 
	  }
	  \multido{\nX=0.0+1.0, \iN=1+1}{8}{
	    \psdot[dotsize=4pt](\nX,0)
	    \rput(\nX,-0.4){\iN}
	  }
	  
	  \multido{\nX=1.5+4.0, \iN=1+1}{2}{
	    \Tens[ \nX, 2, 1.0, 0.45, gray, {$\Lambda^{\![1,\iN]}$} ]
	  }
	  \multido{\nX=0.5+2.0}{4}{ \psdot[dotsize=6pt,linecolor=gray](\nX,1)}
	  
      \TopTens[ 3.5, 4, 2.0, 0.3, gray, {$\Lambda^{\![0,1]}$} ]
	  \multido{\nX=1.5+4.0}{2}{ \psdot[dotsize=6pt,linecolor=gray](\nX,3)}	  
	  \uput[180](2.2,3.5){$m$}
	  \uput[0](4.8,3.5){$m$}
	  \psline[arrowsize=0.2, arrowinset=0.15]{->}(-0.3,4.2)(-0.3,2) \uput[0](-0.3,3.2){$L$ layers}
	  \psline[arrowsize=0.2, arrowinset=0.15]{->}(-0.3,-0.9)(6,-0.9) \uput[270](3,-0.9){$N$ sites}
\end{pspicture}
	\def\Rect[#1,#2,#3,#4]{
  \psframe[framearc=0.2, fillstyle=solid, fillcolor=#4](!#1 #3 sub #2 #3 sub)(!#1 #3 add #2 #3 add)
}

\def\Shadow[#1,#2,#3]{
  \psframe[framearc=0.2, fillstyle=solid, linecolor=shadowgray, fillcolor=shadowgray]
     (!#1 #3 sub #2 #3 sub)(!#1 #3 add #2 #3 add)
}

\psset{arrowinset=0}

\begin{pspicture}(-1.4,-2.4)(7.2,2.8)
	\uput[-90](-1.0,2.8){ (b) }

	\rput(3.6,0){
	\rput(0.05,-0.05){
		\multido{\nAngle=45+90}{4}{
		  \rput{\nAngle}(1.8;\nAngle){ \Shadow[0,0,0.16] }  
		}
		\multido{\nAngle=0+180}{2}{
		  \rput{\nAngle}(0.7;\nAngle){ \Shadow[0,0,0.18] }
		}
		\Shadow[0,0,0.2]
	}

	\rput(0,0){
		\pscircle[linestyle=dashed](0,0){2.4}

		\multido{\nAngle=45+90}{4}{
		  \rput{\nAngle}(0;\nAngle){
			\psline(2.4;-22.5)(1.8;0)
			\psline(2.4;22.5)(1.8;0)
			\psline(1.8;0)(1.4;0)
			\rput(1.8;0){\Rect[0,0,0.16,gray]}    
		  }  
		}
		\multido{\nAngle=0+180}{2}{
		  \rput{\nAngle}(0;\nAngle){
			\psline(1.4;-45)(0.7;0)
			\psline(1.4;45)(0.7;0)
			\psline(0.7;0)(0.35;0)
			\rput(0.7;0){\Rect[0,0,0.18,gray]}
		  }
		}
		\psline(-0.35,0)(0.35,0)
		\Rect[0,0,0.2,gray]

		\multido{\nAngle=112.5+45, \iNum=1+1}{8}{
		  \psdot[dotsize=4pt](2.4;\nAngle)
		  \rput(2.7;\nAngle){\iNum}
		}
		\multido{\nAngle=45+90}{4}{
		  \psdot[dotsize=5pt, linecolor=gray](1.4;\nAngle)
		}
		\psdots[dotsize=5pt, linecolor=gray](-0.35,0)(0.35,0)
	}
	}
\end{pspicture}
	\caption{\label{fig:ttn_structure}
	(a) Structure of the binary TTN with $L$ layers and $N$ sites. The maximal bond dimension is $m$.
	(b) The same TTN displayed in a PBC configuration, showing its natural capability of treating OBC
	and PBC on the same footing.}
\end{figure}

\section{State Architecture and Algorithm}\label{sec:algo}

We consider a one-dimensional lattice with $N=2^L$ sites, where each site has a local Hilbert dimension of $d$.
Our ansatz to approximate a many-body quantum state $|\Psi\rangle = \sum_{\{\vec{\chi}\}}
\Psi_{\chi_1\ldots\chi_N} \bigotimes_{i=1}^{N} |\chi_i\rangle$ on such lattice, where the strings
$\vec{\chi}=(\chi_1\ldots\chi_N)$ label the configurations of the $N$ sites in some local ``canonical'' basis
(with $\chi_i=1\ldots d$), is displayed in Fig.~\ref{fig:ttn_structure}: 
It is a binary TTN~\cite{VidalTTN,PsiTTN2010,TagliacozTTN},
a hierarchical structure consisting of $L$ layers of tensors $\Lambda^{[l,n]}$ with three indices each, where
$l=0\ldots L-1$ indicates the layer and $n=1\ldots 2^l$ denotes the horizontal position of the tensor.
The sketch follows the usual convention of drawing tensor indices as ``legs'' or ``links'';
joining two tensor legs has the usual meaning of a contraction, i.e.~a summation over the corresponding indices
of the tensor elements product.
All tensors in Fig.~\ref{fig:ttn_structure} have three legs, except for the top tensor
$\Lambda^{[0,1]}$ which is two-legged:  
it can be viewed as the contraction of a three-legged tensor 
with a vector encompassing a wavefunction on a renormalized (degenerate) manifold~\cite{Aguado}. 
The physical sites of the chain are represented by the dots attached to the bottom of the lowest
layer of tensors. Each tensor effectively merges two sites into a single ``virtual'' site, allowing one to interpret the
tensors as coarse-graining linear maps in a RG flow. Labeling the sites
(either physical or virtual) with $[l,n]$
($l=1\ldots L$: layer coordinate, $n=1\ldots 2^l$: horizontal coordinate), the tensor
$\Lambda^{[l,n]}$ maps the two sites $[l+1,2n-1]$ and $[l+1,2n]$ to the site $[l,n]$.
Consequently, the full Hilbert spaces of the sites in layer $l$ have dimension $M(l) = d^{\,2^{L-l}}$.
Such dimension is exponentially growing in the number $2^{L-l}$ of physical sites blocked together in layer $l$, and
therefore a numerically efficient representation of such degrees of freedom
requires some kind of space truncation.
The most easily controlled truncation method is fixing a maximally allowed value $m$
(an upper bound to the so-called ``bond dimension''), resulting in $M(l)=\min(d^{\,2^{L-l}}, m)$.
Moreover, the total number of tensors in the binary tree network is $N-1$,
and since we are not introducing additional constraints in the variational picture,
the total number of parameters in the TTN representation ultimately scales like $\mathcal{O}(N \, m^3)$.

In this manuscript, we are going to apply this variational ansatz 
to approximate the ground state of nearest-neighbor interacting spin-Hamiltonians on a lattice.
The application to bosons is straightforward, while extension to 
fermions is carried out via standard Jordan-Wigner transformation~\cite{Batista2001}.
For convenience, we write the local fields and the spin-spin interactions separately,
resulting in a Hamiltonian of the following form:
\begin{equation}
  H = \sum_{n} \cH^{[n]} +
      \sum_{n} \;
      \sum_\alpha \lambda_\alpha \, \cHl^{\alpha [n]} 
      \cHr^{\alpha [n+1]} \; ,
  \label{eq:ttnham}
\end{equation}
where $\cH^{[n]}$ (local term), $\cHl^{[n]}$ (left interaction term) $\cHr^{[n]}$ (right  
interaction term) are on-site operators acting on site $n$.
The index $\alpha$ accounts for the fact that the interaction term can consist of several tensor-product contributions,
weighted by their couplings $\lambda_\alpha$.
A strong point of the TTN ansatz is that it adapts comparatively well to both the
OBC ($n=1\ldots N-1$) setting and the PBC  ($n=1\ldots N$ and $N+1\equiv 1$) setting.
In the two cases, the computational costs are equal, and the numerical precisions compatible.

\begin{figure}
	\begin{pspicture}(0,-1)(8,1.4)
	  \Tens[0.5, 0.0, 0.5, 0.55, gray, {$\Lambda\;$}]
 	  \rput(1.7,0.0){$=$}
 	  \uput[d](1.7,0.0){QR dec.}
 	  \IsoTens[2.9, 0.0, 0.5, 0.55, 3, lightgray, {$Q\;$}]    
 	  \rput{-45}(2.9,0.5){\Op[0, 0.5, 0.5, 0.55, lightgray, {$R\;$}]}

	  \uput[0](3.7,0.0){, with}
 	  \IsoTens[5.5, 0.6, 0.5, 0.55, 3, lightgray, {$Q\;$}]
 	  \IsoTensDag[5.5, -0.6, 0.5, 0.55, 3, lightgray, {$Q^*$}]
 	  \psline(5,-0.1)(5,0.1)
 	  \psline(6,-0.1)(6,0.1)
 	  
 	  \rput(6.3,0){$=$}
 	  \Op[6.9, 0, 0.5, 0.55, lightgray, {$\openone\;$}]
		
      \rput(7.5,0){$=$}
      \psline(7.9,-0.5)(7.9,0.5)
\end{pspicture}
	\caption{\label{fig:tensisom}Isometrization of a three-legged tensor $\Lambda$. The arrow indicates
	the leg with respect to which the tensor is isometric.
	In the picture, $Q^*$ is the element-wise complex complex conjugate of $Q$, and it also
	drawn as vertically reflected.}
\end{figure}
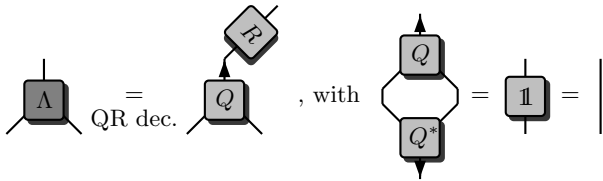

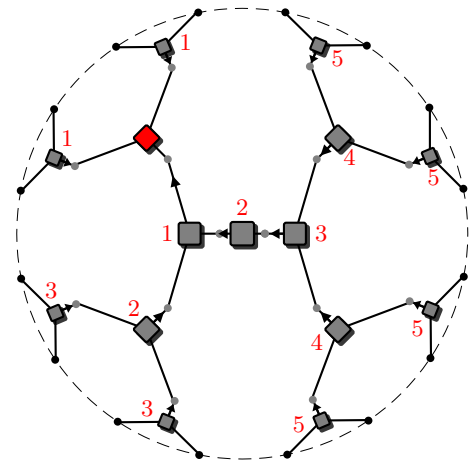
\begin{figure}
	\def\Rect[#1,#2,#3,#4]{
  \psframe[framearc=0.2, fillstyle=solid, fillcolor=#4](!#1 #3 sub #2 #3 sub)(!#1 #3 add #2 #3 add)
}

\def\Shadow[#1,#2,#3]{
  \psframe[framearc=0.2, fillstyle=solid, linecolor=shadowgray, fillcolor=shadowgray]
     (!#1 #3 sub #2 #3 sub)(!#1 #3 add #2 #3 add)
}

\psset{arrowinset=0}

\begin{pspicture}(-3,-3)(3,3)
	\rput(0.04,-0.04){
		\multido{\nAngle=22.5+45}{8}{
		  \rput{\nAngle}(2.7;\nAngle){ \Shadow[0,0,0.10] }
		}
		\multido{\nAngle=45+90}{4}{
		  \rput{\nAngle}(1.8;\nAngle){ \Shadow[0,0,0.14] }  
		}
		\multido{\nAngle=0+180}{2}{
		  \rput{\nAngle}(0.7;\nAngle){ \Shadow[0,0,0.16] }
		}
		\Shadow[0,0,0.17]
	}

	\rput(0,0){
		\pscircle[linestyle=dashed, linewidth=0.1pt](0,0){3}

		\multido{\nAngle=22.5+45}{8}{
		  \rput{\nAngle}(0;\nAngle){
			\psline(3;-11.25)(2.7;0)
			\psline(3;11.25)(2.7;0)
			\psline(2.7;0)(2.4;0)
			\psline[arrowsize=0.13 0, arrowlength=0.7]{->}(2.7;0)(2.45;0)
			\rput(2.7;0){\Rect[0,0,0.10,gray]}        
		  }
		}
		\multido{\nAngle=225+90}{3}{
		  \rput{\nAngle}(0;\nAngle){
			\psline(2.4;-22.5)(1.8;0)
			\psline(2.4;22.5)(1.8;0)
			\psline(1.8;0)(1.4;0)
			\psline[arrowsize=0.14 0, arrowlength=1.0]{->}(1.8;0)(1.47;0)		
			\rput(1.8;0){\Rect[0,0,0.14,gray]}    
		  }  
		}
		\multido{\nAngle=135+90}{1}{
		  \rput{\nAngle}(0;\nAngle){
			\psline(2.4;-22.5)(1.8;0)
			\psline(2.4;22.5)(1.8;0)
			\psline(1.8;0)(1.4;0)
			\rput(1.8;0){\Rect[0,0,0.14,red]}    
		  }  
		}	
		\multido{\nAngle=0+180}{1}{
		  \rput{\nAngle}(0;\nAngle){
			\psline(1.4;-45)(0.7;0)
			\psline(1.4;45)(0.7;0)
			\psline(0.7;0)(0.3;0)
			\psline[arrowsize=0.14 0, arrowlength=0.9]{->}(0.7;0)(0.37;0)		
			\rput(0.7;0){\Rect[0,0,0.16,gray]}
		  }
		  \psdot[dotsize=3pt, linecolor=gray](0.3;\nAngle)
		}
		\multido{\nAngle=180+0}{1}{
		  \rput{\nAngle}(0;\nAngle){
			\psline(1.4;-45)(0.7;0)
			\psline(1.4;45)(0.7;0)
			\psline(0.7;0)(0.3;0)
			\rput(0.7;0){\psline[arrowsize=0.14 0, arrowlength=1.0]{->}(0.7;-73.7)(0.8;-73.7)}
			\rput(0.7;0){\Rect[0,0,0.16,gray]}
		  }
		  \psdot[dotsize=3pt, linecolor=gray](0.3;\nAngle)
		}	
		\psline(-0.3,0)(0.3,0)

		\multido{\nAngle=101.25+22.5, \iNum=1+1}{16}{
		  \psdot[dotsize=3pt](3;\nAngle)
		}
		\multido{\nAngle=22.5+45}{8}{
		  \psdot[dotsize=3pt, linecolor=gray](2.4;\nAngle)
		}
		\multido{\nAngle=45+90}{4}{
		  \psdot[dotsize=3pt, linecolor=gray](1.4;\nAngle)
		}
		\psdots[dotsize=3pt, linecolor=gray](-0.3,0)(0.3,0)
	
		\Rect[0,0,0.17,gray]
		\psline[arrowsize=0.14 0, arrowlength=0.9]{->}(-0.2,0)(-0.32,0)
		
		\rput(2.65;106.5){\red $1$}
		\rput(2.65;151.5){\red $1$}
		\rput(2.65;196.5){\red $3$}
		\rput(2.65;241.5){\red $3$}
		\rput(2.65;286.5){\red $5$}
		\rput(2.65;331.5){\red $5$}
		\rput(2.65;16.5){\red $5$}
		\rput(2.65;61.5){\red $5$}
		
		\rput(1.75;214.5){\red $2$}
		\rput(1.75;304.5){\red $4$}
		\rput(1.75;35.5){\red $4$}
		
		\rput(1.05;0){\red $3$}
		\rput(1.0;180){\red $1$}
		
		\rput(0.35;90){\red $2$}		
	}
\end{pspicture}
	\caption{\label{fig:tensisomdirect}TTN isometrization with respect to the tensor highlighted
	in red, \ie all isometrization arrows point towards this tensor. To obtain this gauge, 
	QR-decompose all tensors in order of decreasing distance (indicated by red numbers).}
\end{figure}

A key requirement for any tensor network representation is its efficient contractibility,
which is instrumental to gain access to physically sensible information on the quantum many-body state,
such as its energy or expectation values of observables.
Indeed, a prominent advantage of TTN architectures is that they are
algebraically contractible, thus providing exact expectation values efficiently, without
the need of stochastic sampling of their variational data~\cite{Contractible1, Contractible2}.
One can easily identify two properties of TTN that make this possible:
The first is the loop-less structure of the tree network 
and the second is the exploitation of a flexible, adaptive 
isometric gauge selection. 
The latter is based on a straightforward generalization of the QR-decomposition~\cite{Horn2012} applied to three-legged
tensors $\Lambda$, which produces a directed-isometric tensor $Q$ and a matrix $R$, such that
\begin{equation}
  \Lambda_{\alpha_1 \alpha_2 \alpha_3} = Q_{\alpha_1 \alpha_2 \beta_1} R_{\beta_1 \alpha_3}\; ,
  \quad Q^*_{\alpha_1 \alpha_2 \beta_1} Q_{\alpha_1 \alpha_2 \beta_2} = \delta_{\beta_1 \beta_2}\; ,
  \label{eq:tensisom} 
\end{equation}
where $Q$ is isometric with respect to the third leg (in the graphical notation of the TTN, we draw an
outgoing arrow from $Q$ on that leg).
In Fig.~\ref{fig:tensisom} we report Eq.~\eqref{eq:tensisom}
expressed in graphical notation.
%%%%%%%%%%%%%%%%%%%%%%%%%%%%%%%%%%%%%%%%%%%%%%%%%%
%%%%%%%%%%%%%%%%%%%%%%%%%%%%%%%%%%%%%%%%%%%%%%%%%%
We say that the TTN is isometrized with respect to tensor $\Lambda^{[l,n]}$ if all the other tensors
in the tree are isometrized in the direction of this tensor according to the network structure,
i.e.~all the arrows are pointing towards
$\Lambda^{[l,n]}$. Note that due to the loopless topology of the TTN such an isometric gauge is
always unique (apart from unitary gauge transformations) 
and efficiently attainable: just perform the QR-decomposition Eq.~\eqref{eq:tensisom} for all the tensors,
starting from the most distant ones (in the network metric,
\ie the graph distance of two nodes in the tree network) 
and absorbing the gauge matrices $R$ into tensors yet to be isometrized, 
until reaching $\Lambda^{[l,n]}$. This concept is illustrated in 
Fig.~\ref{fig:tensisomdirect}.
Also, note that in the adaptive isometric gauge the calculation of the many-body state norm 
$\langle \Psi | \Psi \rangle$ simply collapses to
${\Lambda^*}^{[l,n]}_{\alpha_1 \alpha_2 \alpha_3} \Lambda^{[l,n]}_{\alpha_1\alpha_2\alpha_3}$ 
regardless of the node position in the network, 
since all the other partial contractions cancel to identities due to the isometry condition.

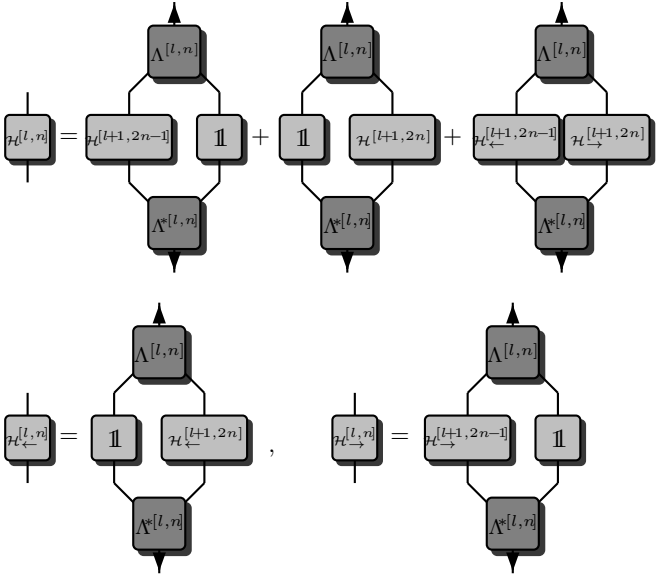
\begin{figure}
	\begin{pspicture}(0,-5.4)(\plotwidth,1.8)
	  \Op[0.35, 0, 0.6, 0.52, lightgray, \tiny{$\cH^{\![l,n\!]}\;\,$}]  
	  \rput(0.9,0){$=$}
	  \rput(2.3,0){
 	    \IsoTens[0, 1.2, 0.6, 0.6, 3, gray, \scriptsize{$\Lambda^{[l,n]}\;$}]
 	    \IsoTensDag[0, -1.2, 0.6, 0.6, 3, gray, \scriptsize{${\Lambda^{\!\!*}}^{\![l,n\!]}\;$}]
 	    \OpRect[-0.6, 0, 0.6, 0.52, lightgray, \tiny{$\cH^{\![l\!+\!1,2n\!-\!1\!]}\;\;$}]
 	    \Op[0.6, 0, 0.6, 0.52, lightgray, \tiny{$\openone\;$}]
 	  }
 	  \rput(3.45,0){$+$}
 	  \rput(4.6,0){
 	    \IsoTens[0, 1.2, 0.6, 0.6, 3, gray, \footnotesize{$\Lambda^{[l,n]}\;$}]
 	    \IsoTensDag[0, -1.2, 0.6, 0.6, 3, gray, \footnotesize{${\Lambda^{\!\!*}}^{\![l,n\!]}\;$}]
 	    \Op[-0.6, 0, 0.6, 0.52, lightgray, \tiny{$\openone\;$}]
 	    \OpRect[0.6, 0, 0.6, 0.52, lightgray, \tiny{$\cH^{[l\!+\!1,2n]}\;$}]
 	  }
 	  \rput(6.0,0){$+$}
 	  \rput(7.45,0){
 	    \IsoTens[0, 1.2, 0.6, 0.6, 3, gray, \footnotesize{$\Lambda^{[l,n]}\;$}]
 	    \IsoTensDag[0, -1.2, 0.6, 0.6, 3, gray, \footnotesize{${\Lambda^{\!\!*}}^{\![l,n\!]}\;$}]
 	    \OpRect[-0.6, 0, 0.6, 0.52, lightgray, \tiny{$\cHl^{\![l\!+\!1,2n\!-\!1\!]}\;\;$}]
 	    \OpRect[0.6, 0, 0.6, 0.52, lightgray, \tiny{$\cHr^{[l\!+\!1,2n]}\;$}]
 	  }
 	  
  	  \Op[0.35, -4, 0.6, 0.52, lightgray, \tiny{$\cHl^{\![l,n\!]}\;\,$}]   	  
 	  \rput(0.9,-4){$=$}
 	  \rput(2.1,-4){
 	    \IsoTens[0, 1.2, 0.6, 0.6, 3, gray, \footnotesize{$\Lambda^{[l,n]}\;$}]
 	    \IsoTensDag[0, -1.2, 0.6, 0.6, 3, gray, \footnotesize{${\Lambda^{\!\!*}}^{\![l,n\!]}\;$}]
 	    \Op[-0.6, 0, 0.6, 0.52, lightgray, \tiny{$\openone\;$}]
 	    \OpRect[0.6, 0, 0.6, 0.52, lightgray, \tiny{$\cHl^{[l\!+\!1,2n]}\;$}]
 	  }
 	  \rput(3.6,-4.2){$,$}
 	  
  	  \Op[4.7, -4, 0.6, 0.52, lightgray, \tiny{$\cHr^{\![l,n\!]}\;\,$}]   	  
 	  \rput(5.3,-4){$=$}
 	  \rput(6.8,-4){
 	    \IsoTens[0, 1.2, 0.6, 0.6, 3, gray, \footnotesize{$\Lambda^{[l,n]}\;$}]
 	    \IsoTensDag[0, -1.2, 0.6, 0.6, 3, gray, \footnotesize{${\Lambda^{\!\!*}}^{\![l,n\!]}\;$}]
 	    \OpRect[-0.6, 0, 0.6, 0.52, lightgray, \tiny{$\cHr^{\![l\!+\!1,2n\!-\!1\!]}\;\;$}]
 	    \Op[0.6, 0, 0.6, 0.52, lightgray, \tiny{$\openone\;$}]
 	  }
\end{pspicture}
	\caption{\label{fig:tens_map}Mapping operation induced by the tensor $\Lambda^{[l,n]}$. The
	operation maps the effective Hamiltonian terms at sites [l+1,2n-1] and [l+1, 2n]
	to the new terms at site $[l,n]$. The index $\alpha$ has been dropped for simplicity.}
\end{figure}

\begin{figure}
	\begin{pspicture}(0,-5.1)(\plotwidth,1.8)
	  \rput[l](0,0){$H^{[l,n]}_{\text{eff}} \; |\Lambda^{[l,n]}\rangle =$}
	  \Tens[3, 0, 0.6, 0.65, red, {$\Lambda^{[l,n]}\,$}]
	  \OpRect[2.4, -1.2, 0.6, 0.52, lightgray, \tiny{$\cH^{\![l\!+\!1,2n\!-\!1\!]}\;\;$}]	    
 	  \rput(4.25,0){$+$}
 	  \Tens[5.5, 0, 0.6, 0.65, red, {$\Lambda^{[l,n]}\,$}]
 	  \OpRect[6.1, -1.2, 0.6, 0.52, lightgray, \tiny{$\cH^{[l\!+\!1,2n]}\;$}]
 	  \rput(6.75,0){$+$}
 	  \Tens[8, 0, 0.6, 0.65, red, {$\Lambda^{[l,n]}\,$}]
 	  \OpRect[8, 1.2, 0.6, 0.52, lightgray, \tiny{$\cH^{[l,n]}\;$}]

	  \uput[0](1.5,-3.8){$+$}
	  \Tens[3, -3.8, 0.6, 0.65, red, {$\Lambda^{[l,n]}\,$}] 
	  \OpRect[2.4, -5, 0.6, 0.52, lightgray, \tiny{$\cHl^{\![l\!+\!1,2n\!-\!1\!]}\;\;$}] 
	  \OpRect[3.6, -5, 0.6, 0.52, lightgray, \tiny{$\cHr^{[l\!+\!1,2n]}\;$}]
	  \rput(4.25,-3.8){$+$}
	  \Tens[5.5, -3.8, 0.6, 0.65, red, {$\Lambda^{[l,n]}\,$}]
	  \OpRect[5.5, -2.6, 0.6, 0.52, lightgray, \tiny{$\cHr^{[l,n]}\;$}]
	  \OpRect[6.1, -5, 0.6, 0.52, lightgray, \tiny{$\cHl^{[l\!+\!1,2n]}\;$}]
	  \rput(6.75,-3.8){$+$}
	  \Tens[8, -3.8, 0.6, 0.65, red, {$\Lambda^{[l,n]}\,$}]
	  \OpRect[8, -2.6, 0.6, 0.52, lightgray, \tiny{$\cHl^{[l,n]}\;$}] 
	  \OpRect[7.4, -5, 0.6, 0.52, lightgray, \tiny{$\cHr^{\![l\!+\!1,2n\!-\!1\!]}\;\;$}]
\end{pspicture}
	\caption{\label{fig:action_effham}Definition of the action of the effective Hamiltonian
	$H^{[l,n]}_{\text{eff}}$ on the degrees of freedom of the tensor $\Lambda^{[l,n]}$. (Again, the
	index $\alpha$ has been dropped for simplicity.)}
\end{figure}
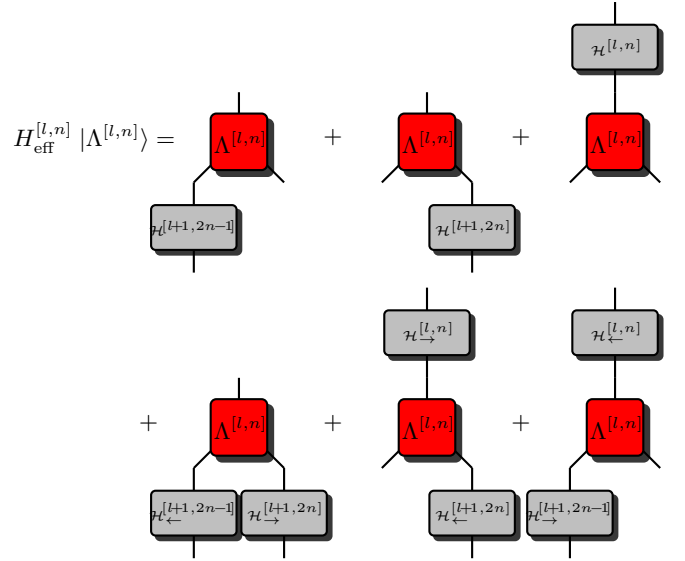

Such a flexible gauge selection provides a crucial advantage
along the search for the best TTN representation of the ground state of the Hamiltonian  $H$ in Eq.~\eqref{eq:ttnham}.
Expressed in a variational sense, the task consists in searching the set of
tensors $\{ \Lambda^{[l,n]} \}$ such 
that $E = \langle \Psi | H |  \Psi \rangle / \langle \Psi | \Psi \rangle$ is minimal.
For practical values of $N$ and $m$ the complete space of variational parameters of all tensors
combined is too large to allow a successful application of a direct search optimization; instead, the
approach pursued here relies on an iterative strategy,
optimizing one tensor at a time while assuming all other tensors to be fixed.
A sensible reason to choose this strategy is that the optimization problem for a single tensor
$\Lambda^{[l,n]}$ actually reduces to a simple eigenvalue problem for an effective Hamiltonian
$H^{[l,n]}_{\text{eff}}$ acting on the degrees of freedom of $\Lambda^{[l,n]}$ alone, 
in an analogous fashion to the density matrix RG with single center site\cite{Whiteonesite}. 
To see this,
we define iteratively for each virtual site $[l,n]$ the effective Hamiltonian terms $\cH^{[l,n]}$,
$\cHl^{\alpha [l,n]}$ and $\cHr^{\alpha [l,n]}$, resulting from performing the isometric mapping operation
sketched in Fig.~\ref{fig:tens_map}. By identifying the effective Hamiltonian terms on the physical sites layer
as the original Hamiltonian, i.e.~$\cH^{[L,n]} \equiv \cH^{[n]}$, $\cHlr^{\alpha [L,n]} \equiv \cHlr^{\alpha [n]}$ 
the mapping is a well-defined operation for every link.
%%%%%%%%%%%%%%%%%%%%%%%%%%%%%%%%%%%%%%%%%%%%%%%%%%
%%%%%%%%%%%%%%%%%%%%%%%%%%%%%%%%%%%%%%%%%%%%%%%%%%
On the other hand, we do not have to explicitly define effective identity operators $\mathcal{N}^{[l,n]}$,
since the identity operator~$\openone$ is invariant under the mapping by construction.
The mapping is always to be carried in the direction of the adaptive gauge isometrization: starting from
contracting the Hamiltonian pieces which are farther from the node $[l,n]$ in the network metric,
and then proceeding closer and closer, until the effective Hamiltonian only acts on $\Lambda^{[l,n]}$ itself.
According to this picture, the calculation of the energy expectation value for the TTN
can be written down as $E=\langle
\Lambda^{[l,n]} | H^{[l,n]}_{\text{eff}} | \Lambda^{[l,n]} \rangle$, where the action of
$H^{[l,n]}_{\text{eff}}$ on $|\Lambda^{[l,n]}\rangle$ is meant as indicated in
Fig.~\ref{fig:action_effham}.
%%%%%%%%%%%%%%%%%%%%%%%%%%%%%%%%%%%%%%%%%%%%%%%%%%
%%%%%%%%%%%%%%%%%%%%%%%%%%%%%%%%%%%%%%%%%%%%%%%%%%
The minimization problem for $\Lambda^{[l,n]}$, subject to the constraint of normalization, is then
easily solved by the method of Lagrange multipliers. We then write the following Lagrangian
\begin{eqnarray}
  \mathcal{L}\left( |\Lambda^{[l,n]} \rangle, \langle \Lambda^{[l,n]} |, \lambda \right)
     = &&\langle \Lambda^{[l,n]} | H^{[l,n]}_{\text{eff}} | \Lambda^{[l,n]} \rangle \nonumber\\
      && - \lambda \, \left( \langle \Lambda^{[l,n]} | \openone | \Lambda^{[l,n]} \rangle -1 \right)
      \; ,  \qquad
  \label{eq:lagrangian}
\end{eqnarray}
and since tensors different from $\Lambda^{[l,n]}$ are assumed to be fixed,
the Euler-Lagrange equation simply reads
\begin{equation}
  H^{[l,n]}_{\text{eff}} | \Lambda^{[l,n]} \rangle = \lambda \, \openone |\Lambda^{[l,n]} \rangle
  \; , \quad \langle \Lambda^{[l,n]} | \openone | \Lambda^{[l,n]} \rangle = 1 \, .
  \label{eq:extrcond}
\end{equation}
which is a \emph{standard} eigenvalue problem (SEP) for $H^{[l,n]}_{\text{eff}}$.
By linearity, it is
readily seen that the normalized eigenvector of $H^{[l,n]}_{\text{eff}}$ corresponding to the lowest
eigenvalue is the best choice for $\Lambda^{[l,n]}$ to minimize $E$.
We highlighted on purpose the identity operators in Eq.~\eqref{eq:extrcond},
in order to stress the benefit that is obtained from the
isometric gauge selection:
In fact, choosing a different gauge would require to substitute the~$\openone$ for a (nontrivial)
effective identity operator $\mathcal{N}^{[l,n]}$, turning Eq.~\eqref{eq:extrcond} into a
\emph{generalized} eigenvalue problem, which is significantly more demanding and unstable
than a SEP when addressed numerically~\cite{bai2000templates}.

%%%%%%%%%%%%%%%%%%%%%%%%%%%%%%%%%%%%%%%%%%%%%%%%%%
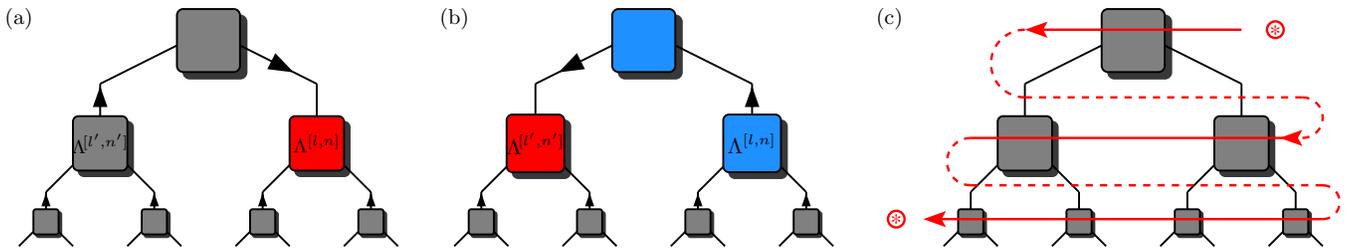
\begin{figure*}
	\psscalebox{0.9}{
	\begin{pspicture}(0,0)(6.2,3.8)
	  \uput[-90](0,3.7){(a)}
	  \IsoTens[0.4,0.4,0.4,0.5,3,gray,{}]
	  \IsoTens[2.0,0.4,0.4,0.5,3,gray,{}]
	  \IsoTens[3.6,0.4,0.4,0.5,3,gray,{}]
	  \IsoTens[5.2,0.4,0.4,0.5,3,gray,{}]
	  \IsoTens[1.2,1.6,0.8,0.52,3,gray,\footnotesize{$\Lambda^{\![l^{\prime},n^{\prime}]}$}]
	  \Tens[4.4,1.6,0.8,0.52,red,\footnotesize{$\Lambda^{[l,n]}$}]
	  \IsoTopTens[2.8,3.2,1.6,0.3,2,gray,{}]
	\end{pspicture}}
	\psscalebox{0.9}{
	\begin{pspicture}(0,0)(6.2,3.8)
	  \uput[-90](0,3.7){(b)}
	  \IsoTens[0.4,0.4,0.4,0.5,3,gray,{}]
	  \IsoTens[2.0,0.4,0.4,0.5,3,gray,{}]
	  \IsoTens[3.6,0.4,0.4,0.5,3,gray,{}]
	  \IsoTens[5.2,0.4,0.4,0.5,3,gray,{}]
	  \Tens[1.2,1.6,0.8,0.55,red,{$\Lambda^{\![l^{\prime},n^{\prime}]}$}]
	  \IsoTens[4.4,1.6,0.8,0.55,3,lightblue,{$\Lambda^{[l,n]}$}]
	  \IsoTopTens[2.8,3.2,1.6,0.3,1,lightblue,{}]
	\end{pspicture}}
	\psscalebox{0.9}{
	\begin{pspicture}(-0.8,0)(5.9,3.8)
	  \uput[-90](-0.8,3.7){(c)}
	  \Tens[0.4,0.4,0.4,0.5,gray,{}]
	  \Tens[2.0,0.4,0.4,0.5,gray,{}]
	  \Tens[3.6,0.4,0.4,0.5,gray,{}]
	  \Tens[5.2,0.4,0.4,0.5,gray,{}]
	  \Tens[1.2,1.6,0.8,0.52,gray,{}]
	  \Tens[4.4,1.6,0.8,0.52,gray,{}]
	  \TopTens[2.8,3.2,1.6,0.3,gray,{}]
	  
	  \psset{arrowsize=0.17 2, arrowinset=0.2, linewidth=1.0pt, linecolor=red, dash=3pt 3pt}
	  \rput(-0.7,0.4){\pscirclebox{\psdot[dotsize=4pt, dotstyle=asterisk](0,0)}}
	  \psline{<-}(-0.3,0.4)(5.6,0.4)
	  \psarc[linestyle=dashed](5.6,0.65){0.25}{-90}{90}
	  \psline[linestyle=dashed](0.4,0.9)(5.6,0.9)
	  \psarc[linestyle=dashed](0.4,1.25){0.35}{90}{270}
	  \psline{-<}(0.4,1.6)(5.3,1.6)
	  \psarc[linestyle=dashed](5.3,1.9){0.3}{-90}{90}
	  \psline[linestyle=dashed](1.2,2.2)(5.3,2.2)
	  \psarc[linestyle=dashed](1.2,2.7){0.5}{90}{270}
	  \psline{<-}(1.2,3.2)(4.4,3.2)
	  \rput(4.9,3.2){\pscirclebox{\psdot[dotsize=4pt, dotstyle=asterisk](0,0)}}
	\end{pspicture}}
	\caption{\label{fig:ttn_opt}
	Generic optimization move: (a) After optimizing $\Lambda^{[l,n]}$, 
	the tensor $\Lambda^{[l^\prime,n^\prime]}$ is targeted for optimization. 
	(b) Only tensors and effective Hamiltonian terms located on the path connecting $\Lambda^{[l,n]}$
	and $\Lambda^{[l^\prime,n^\prime]}$ (colored in blue) need to be updated in order to enable the 
	optimization of $\Lambda^{[l^\prime,n^\prime]}$. 
	(c) Targeting each tensor in the tree
	multiple times results in a sweeping pattern. After completing a sweep, resume at the top 
	(as indicated by the encircled marks).}
\end{figure*}
%%%%%%%%%%%%%%%%%%%%%%%%%%%%%%%%%%%%%%%%%%%%%%%%%%

After these considerations, the ground state search algorithm is summarized as follows:
\begin{enumerate}[(i)]
  \item Initialize all tensor entries:
  \begin{enumerate}[(a)]
	\item by picking a random state in the TTN manifold;
	\item by selecting a particularly symmetric or meaningful state (e.g. a ferromagnetic product state);
	\item by performing some iterations of an exponentially growing DMRG-like procedure;
  \end{enumerate}
  In the following, we focus on strategy (a) to prove that the algorithm is robust, as
  its convergence is ultimately insensitive to the initialization. 
  \item Select a tensor $\Lambda^{[l,n]}$ in the network.
    Isometrize the TTN in the direction of $\Lambda^{[l,n]}$ and perform the mapping
    operations according to the directed network. Optimize $\Lambda^{[l,n]}$ by solving
    Eq.~\eqref{eq:extrcond}.
  \item From $\Lambda^{[l,n]}$ move to the next tensor $\Lambda^{[l^\prime, n^\prime]}$, adjusting
    the isometrization and updating the effective Hamiltonian terms. Note that only
    tensors located on the path connecting $\Lambda^{[l,n]}$ and $\Lambda^{[l^\prime,n^\prime]}$ are
    affected by this move (see Fig.~\ref{fig:ttn_opt}). Having determined the new effective
    Hamiltonian, optimize $\Lambda^{[l^\prime, n^\prime]}$ via Eq.~\eqref{eq:extrcond} again.
  \item Repeat (iii), targeting each tensor in the tree by following some ``sweeping'' pattern 
    (\eg the one indicated in Fig.~\ref{fig:ttn_opt}); stop when convergence in the ground state
    energy is reached, according to some precision threshold.
\end{enumerate}

The sweeping action is the key point which pushes the TTN representation beyond the simple numerical
real space RG-flow: Optimizing multiple times the same tensor while tuning the environment 
(i.e. the rest of the network surrounding it) 
makes the algorithm a complete variational approach, guaranteed to converge in the TTN manifold.
Summing up, we have reduced the energy minimization problem to a sequence of QR-decompositions,
linear mapping operations and SEPs, all of which can be carried out in a numerically stable fashion with a
computational cost of $\mathcal{O}(m^4)$. Let us stress that this scaling behavior is independent of
the boundary conditions chosen, at a difference with MPS ans\"atze;  
the only change accompanying a switch from PBC to OBC consists in
omitting the terms that mediate the interaction between the physical sites $1$ and~$N$ when
calculating the action of the effective Hamiltonian (Fig.~\ref{fig:action_effham}).
These terms only occur for the outermost tensors in a layer (i.e.~$n=1$ or $n=2^l$),
and thus produce a subleading change in the overall computational cost.
Moreover, we remark that the $\mathcal{O}(m^4)$ contraction cost relies only on the loop-free network structure and
the fact that tensors have three legs: this means that 
the binary TTN ansatz can naturally be extended to other
lattices and dimensionalities (e.g.~2D, Cayley trees~\cite{Verstraetedenrimer})
without increasing the numerical effort for the contractions.

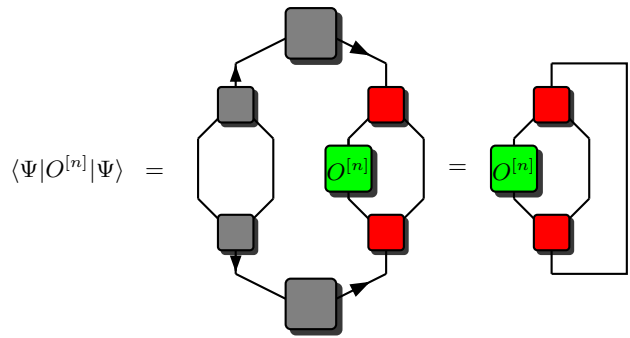
\begin{figure}
  \begin{pspicture}(0,-2)(8.2,2)
    \rput[l](0,0){$\langle\Psi| O^{[n]} | \Psi\rangle \;\;=$}
    \IsoTens[3,0.9,0.5,0.5,3,gray,{}]
    \Tens[5,0.9,0.5,0.5,red,{}]
    \IsoTopTens[4,1.9,1,0.35,2,gray,{}]
    \IsoTensDag[3,-0.9,0.5,0.5,3,gray,{}]
    \TensDag[5,-0.9,0.5,0.5,red,{}]
    \IsoTopTensDag[4,-1.9,1,0.35,1,gray,{}]

    \multido{\nX=2.5+1.0}{4}{
      \psline(\nX,0.4)(\nX,-0.4)
    }
	\Op[4.5,0,0.4,0.8,green,$O^{[n]}$]
    \rput(5.95,0){$=$}
	\rput[l](7.2,0){
      \Tens[0,0.9,0.5,0.5,red,{}]
      \TensDag[0,-0.9,0.5,0.5,red,{}]    
      \Op[-0.5,0,0.4,0.8,green,$O^{[n]}$]
      \psline(0.5,0.4)(0.5,-0.4)
      \psline(0,1.4)(1,1.4)(1,-1.4)(0,-1.4)
    }
  \end{pspicture}
  \caption{\label{fig:ttn_localobs}Expectation value of a local observable $O^{[n]}$. Due to the
  isometry condition, the calculation always reduces to a contraction involving only three tensors.}
\end{figure}

We conclude this section by describing how to extract expectation values of the TTN state
$|\Psi\rangle$. This is particularly convenient for local observables $O^{[n]}$, having support on a single
lattice site $n$. By isometrizing the TTN with respect to $\Lambda^{[L-1, \lceil n/2
\rceil]}$ (i.e.~the tensor directly attached to site~$n$; $\lceil x \rceil$ is the ceiling function
of $x$), the expectation value $\langle \Psi | O^{[n]} | \Psi \rangle$ collapses to a contraction
of only three tensors, as indicated in Fig.~\ref{fig:ttn_localobs}. In the case of two-point
correlators $\langle O^{[n]} O^{[n+r]} \rangle$ (two local observables separated by a distance $r$) 
a similar procedure can be adopted; the only difference is that now nontrivial contractions
arise for all tensors on 
the network path connecting the two sites $n$ and $n+r$ (which is unique thanks
to the loop-free network structure).
Given that the maximum number of
tensors on this path scales logarithmically in the lattice size $N$, two-point correlators can be computed very
efficiently.
%%%%%%%%%%%%%%%%%%%%%%%%%%%%%%%%%%%%%%%%%%%%%%%%%%
%%%%%%%%%%%%%%%%%%%%%%%%%%%%%%%%%%%%%%%%%%%%%%%%%%

\section{Translational invariance in the TTN architecture} \label{sec:traslinv}

%%%%%%%%%%%%%%%%%%%%%%%%%%%%%
\begin{figure*}
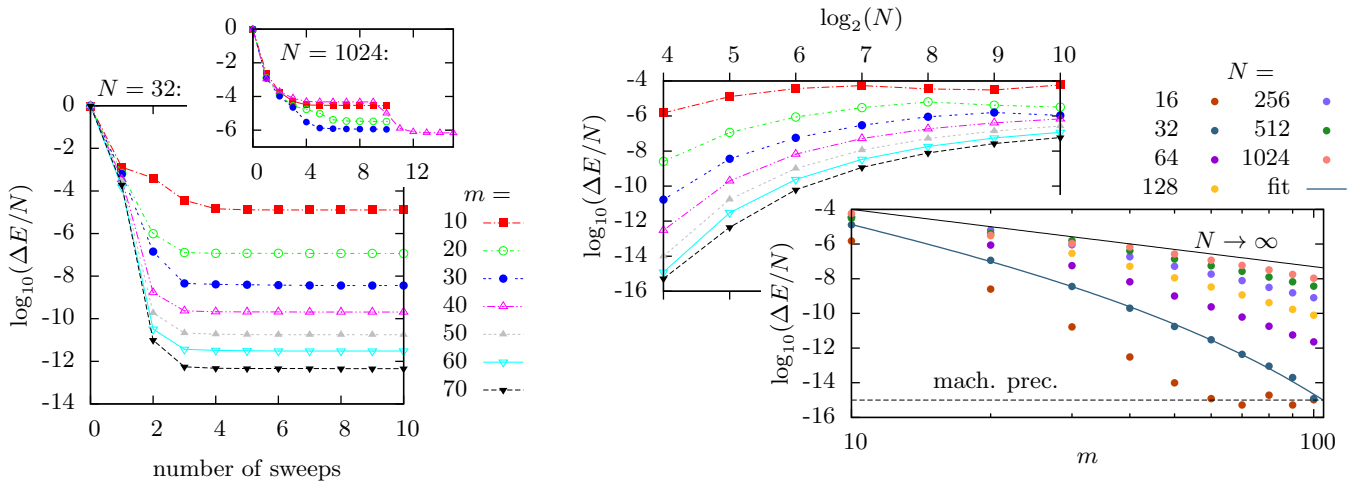

  \begin{minipage}{0.405\textwidth}
    \input{gnuplottex-fig1}
  \end{minipage} \hspace{-3mm}
  \begin{minipage}{0.59\textwidth}
    \input{gnuplottex-fig2}
  \end{minipage}
  \vspace{3mm}
  \caption{\label{fig:ising_energyerr} Ising model: Error of the ground state energy per site
  $\Delta E/N$ as a function of the number of optimization sweeps for two different system sizes $N=32$ and $N=1024$ (left).
  Error as a function of the system size (right-upper). The same quantity as a function of the bond
  dimension $m$. The data for $N=32$ is fitted by $a \cdot m^{-b} \cdot \exp(-c\,m)$, with
  $a=12.8$, $b=5.5$, $c=0.11$. Extrapolation to the thermodynamic limit suggests a polynomial decay
  of the form $\Delta E \propto m^{-3.3}$ (right-lower).}
\end{figure*}
%%%%%%%%%%%%%%%%%%%%%%%%%%%%%%%%

Before benchmarking the presented algorithm, let us analyse the issue of the translational 
invariance of the lattice, which is broken by the design of the TTN ansatz.
Indeed, it is clear from Fig.~\ref{fig:ttn_structure} that some sites 
are better connected with their immediate environment  
(i.e. the neighbouring sites) than others~\cite{Ferris2013}.
For instance, a very poorly connected environment occurs for the sites at $N/2$ and
$N/2+1$, which are nearest neighbors,
and yet they are renormalized together (in the RG-flow picture)
only at the last step, \ie at the very top of the tree network.
The translational symmetry breaking induced by the TTN design makes one wonder
what is the best strategy to obtain most accurate observation results
when we are simulating a translationally invariant model.
A legitimate question is whether it is beneficial or detrimental to 
measure at one site (or region) in particular
rather than translationally averaging the measurements over the lattice.
An analysis in this direction has been recently developed for a two-dimensional TTN design 
and reported in Ref.~\onlinecite{Ferris2013}:
in that scenario, a significantly more
accurate description of local quantities has been obtained by focusing on ``central sites'', which
are defined by the criterion that the Hilbert space of their immediate environment is largest. In our
1D setting we can identify the location~$s_c$ of these sites by alternately following the left (l)
or right (r) branch down the tree, starting from the top, as indicated in the following sketch:
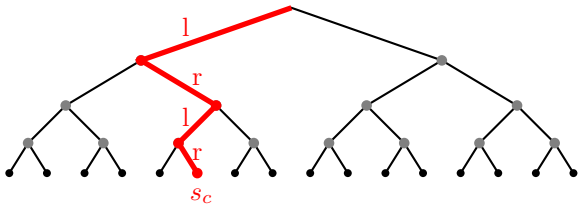
\begin{center}
	  \begin{pspicture}(-4,-0.5)(3.5,2.2)
    \multido{\nA=-4+1.0, \nB=-3.75+1.00, \nC=-3.5+1.0}{8}{ 
       \psline(\nA, 0)(\nB, 0.4)(\nC, 0)
       \psdot[dotsize=3pt,linecolor=gray](\nB,0.4)
    }
    \multido{\nA=-3.75+2.00, \nB=-3.25+2.00, \nC=-2.75+2.00}{4}{ 
       \psline(\nA, 0.4)(\nB, 0.9)(\nC, 0.4) 
    }
    \multido{\nA=-3.25+4.00, \nB=-2.25+4.00, \nC=-1.25+4.00}{2}{ 
       \psline(\nA, 0.9)(\nB, 1.5)(\nC, 0.9) 
    }
    \psline(-2.25,1.5)(-0.25,2.2)(1.75,1.5)
    
    \multido{\nA=-4.0+0.5}{16}{ \psdot[dotsize=3pt](\nA,0) }
    \multido{\nA=-3.75+1.00}{8}{ \psdot[dotsize=4pt, linecolor=gray](\nA,0.4) }
    \multido{\nA=-3.25+2.00}{4}{ \psdot[dotsize=4pt, linecolor=gray](\nA,0.9) }
    \multido{\nA=-2.25+4.00}{2}{ \psdot[dotsize=4pt, linecolor=gray](\nA,1.5) }  
    
    \psline[linecolor=red, linewidth=2pt](-0.25,2.2)(-2.25,1.5)(-1.25,0.9)(-1.75,0.4)(-1.5,0)
    \psdots[dotsize=4pt, linecolor=red](-2.25,1.5)(-1.25,0.9)(-1.75,0.4)(-1.5,0)
    
    \rput(-1.65,1.95){\red l}
    \rput(-1.5,1.25){\red r}
    \rput(-1.65,0.75){\red l}
    \rput(-1.5,0.25){\red r}
    \rput(-1.45,-0.3){\red $s_c$}
  \end{pspicture}
\end{center}
Algebraically, this can be written as
\begin{equation}
 	s_c = 1+\sum_{l=1}^{\lfloor L/2 \rfloor} 2^{L-2l} 
 		= \lfloor (N+2)/3 \rfloor \; ,
 	\label{eq:centsite}
\end{equation}
where $\lfloor x \rfloor$ is the floor function of $x$. 
In order to clarify the situation for the 1D-TTN, we will use a benchmark Ising model to compute
local observables and two-point correlators at all lattice sites and compare them with their
respective system-wide averages. The results of this analysis will be presented 
at the end of the next section.

\section{Benchmarking of the algorithm}

%%%%%%%%%%%%%%%%%%%%%%%%%%%%%%%%%%%%%%%%%%%%%%%%%%
\begin{figure}[t]
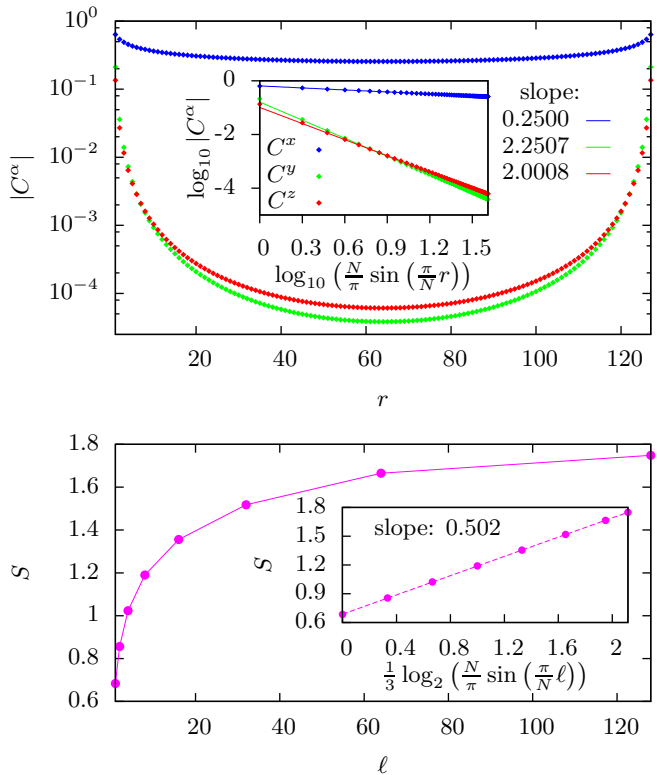

	\input{gnuplottex-fig3}
	\\[3mm]
	\input{gnuplottex-fig4}
	\\[5mm]
	\caption{\label{fig:ising_crit}  (Top) Spin-correlation functions and fitted critical exponents,
	for a PBC quantum Ising chain of $N = 128$ sites, bond dimension $m=100$. 
	The exact exponents are \mbox{$\eta_x=0.25$}, $\eta_y=2.25$, $\eta_z=2$.
	In order to account for the TTN-induced translational symmetry breaking, 
	the correlations of distance $r$ are obtained by averaging over lattice locations 
	within branches wide enough to span~$r$ (see text). 
	(Bottom) Von Neumann entropy of a $N=256$ ($m=100$) chain, for partitions of 
	length $\ell$ and with fitted central charge.
	The exact central charge is $c=0.5$.}
\end{figure}
%%%%%%%%%%%%%%%%%%%%%%%%%%%%%%%%%%%%%%%%%%%%%%%%%%

To test the algorithm outlined in Sec.~\ref{sec:algo} we consider the
spin-$\frac{1}{2}$ Ising model in a transverse field with PBC, defined by the Hamiltonian
\begin{equation}
  H = -\sum_{n=1}^{\text{PBC}} \sigma^x_{n}\sigma^x_{n+1} + h \, \sigma^z_n \; ,
  \label{eq:ham_ising}
\end{equation}
where $\sigma^\alpha_n$ ($\alpha=x,y,z$) is a Pauli matrix acting on spin $n$ and the parameter
$h$ is the external magnetic field. We will focus on the critical point of the model at $|h|=1$,
where, as previously stated and argued e.g. in 
Refs.~\onlinecite{QuMERAchan2008,PsiTTN2010,VidalMERAscale2006}, 
the entanglement scaling of TTN should prove more useful.
This analytically solvable model~\cite{Lieb1961,Pfeuty1970} constitutes a commonly employed
excellent benchmark for the quantities of our interest: namely, 
the ground state energy~$E$, the central charge $c$, and the spin-correlation 
functions $C^\alpha(r) = \langle \sigma^\alpha_n \sigma^\alpha_{n+r} \rangle - 
\langle \sigma^\alpha_n \rangle \langle \sigma^\alpha_{n+r} \rangle$
including their respective critical exponents $\eta_\alpha$.

First of all we report the convergence behavior of the algorithm, shown on the left hand side of
Fig.~\ref{fig:ising_energyerr}: Convergence to the ground state as a function of the number of
optimization sweeps is fast, with roughly fifteen sweeps being sufficient to reliably achieve
the global minimum in the TTN state space, even for big system sizes of $N=1024$ sites. As can be seen
from the upper right plot in Fig.~\ref{fig:ising_energyerr}, the error of the ground state energy
per site tends to be size-independent with increasing $N$, a feature the TTN shares with other
hierarchical tensor network ans\"atze~\cite{Evenbly2009}, when simulating a (1+1)D gapless system.
An intuitive argument to motivate this is that, in order to describe the entanglement of a localized region
with given precision, one has to pick a fixed value~$m$ and not one that scales with the length~$L$:
in a critical scenario, this can happen only if a given~$m$ captures the critical correlation
scaling, \ie the logarithmic entanglement area law violation. 
In the lower right graph we plot the dependency
of the error on the bond dimension $m$. In double-logarithmic representation the curves clearly
exhibit a negative curvature, meaning that the decay is sub-polynomial for any finite $N$; 
in the relevant range of $m$ 
the  behavior can be described well by a combined polynomial-exponential decay of the form $\Delta
E(m) = a \cdot m^{-b} \cdot \exp(-c \, m)$, where $a$, $b$, $c$ are fit parameters dependent on the
system size (see Fig.~\ref{fig:ising_energyerr}). In the thermodynamic limit 
$N \rightarrow \infty$ the decay seems to be governed by a polynomial behavior of the form 
$\Delta E(m) \propto m^{-3.3}$.

In order to assess how the TTN deals with criticality and, consequently, strongly correlated systems,
we report the spin-correlation functions and the von Neumann entropy in Fig.~\ref{fig:ising_crit}.
Since we are at criticality, we expect the correlation functions to decay according to a power law
$C^{\alpha}(r) \propto  \left[ \crd(r) \right]^{-\eta_\alpha}$, where $\crd(r) \equiv N/\pi \cdot
\sin(\pi r / N)$ is the chord function giving the effective distance of two sites at distance $r$
arranged on a ring of $N$ sites~\cite{Caza, Mussardobook, Sachdevbook}.
The critical exponents are analytically known~\cite{Sachdevbook} to be
$\eta_x=1/4$, $\eta_y=9/4$, and $\eta_z=2$.
The via fitting numerically obtained exponents in the upper panel of Fig.~\ref{fig:ising_crit}
show an agreement of the order of $\mathcal{O}(10^{-4})$,
presenting state-of-the-art accuracy to the best of our knowledge~\cite{Evenbly2009}. The von
Neumann entropy $S(\ell) \equiv - \text{Tr}[\rho_\ell \ln \rho_\ell]$
of a partition of $\ell$ sites for PBC critical (1+1)D systems is known to behave according
to $S(\ell) = c/3 \cdot \log_2 \left[ \crd(\ell) \right] + \text{const.}$, where $c$ is the central charge
of the underlying conformal field theory ~\cite{Calabrese2004}, and for the Ising model we have $c=1/2$.
Again, we have very good agreement of the exact value with the numerically determined central charge
$c=0.502$ shown in the lower panel of Fig.~\ref{fig:ising_crit}, proving that the TTN ansatz
is well suited for reproducing a system at criticality.
In particular, we showed that even with a relatively small bond dimension $m$,  
it is perfectly meaningful to simulate critical systems even with hundreds of sites, 
and indeed achieve a high precision of the results.

%%%%%%%%%%%%%%%%%%%%%%%%%%%%%%%%%%%%%%%%%%%%%%%%%%
\begin{figure*}
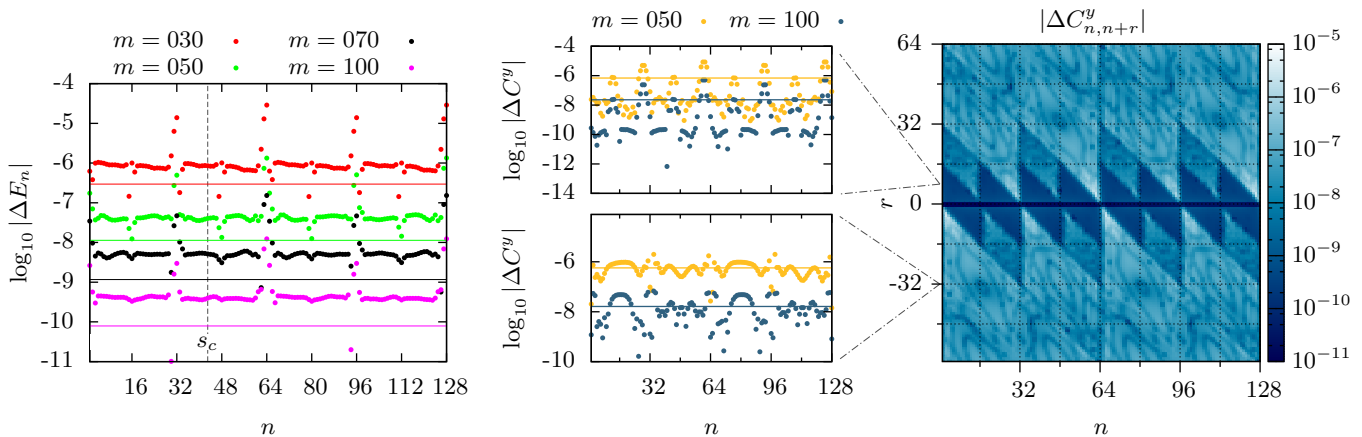

  \begin{minipage}{0.395\textwidth}
    \input{gnuplottex-fig5}
  \end{minipage} \hspace{-3mm}
  \begin{minipage}{0.60\textwidth}
    \input{gnuplottex-fig6}
  \end{minipage} \\[3mm]
  \caption{\label{fig:ising_averaging}Error of the local energy $E_n$ as a function of the lattice
  site $n$ for different bond dimensions $m$, in the critical quantum Ising model with PBC defined  
  in Eq.~\eqref{eq:ham_ising}. The solid lines are the errors of the system-wide 
  averages. The vertical dashed line indicates the location of the center site $s_c$ (left).
  Error of the spin two-point correlator $C^y_{n,n+r}$ for bond dimension $m=100$ (right);
  the two insets show the data at fixed distance $r$ together with system-wide averages for $r=8$
  (top) and $r=-32$ (bottom).}
\end{figure*}
%%%%%%%%%%%%%%%%%%%%%%%%%%%%%%%%%%%%%%%%%%%%%%%%%%

We conclude this section by addressing the question raised in Sec.~\ref{sec:traslinv}, 
\ie whether or not it is beneficial to average over expectation values.
For this purpose we compute the local energy $E_n=-\langle
\sigma^x_n \sigma^x_{n+1} \rangle + \langle \sigma^z_n \rangle$ for the critical Ising model
Eq.~\eqref{eq:ham_ising} and plot the absolute value of its error $\Delta E_n$ as a function of the
site location (see Fig.~\ref{fig:ising_averaging}). Also shown is the error of the system-wide average $\sum_{n=1}^N
E_n / N$. It is clear from Fig.~\ref{fig:ising_averaging} that the averaged quantity is about one order
of magnitude more accurate than the vast majority of the individual measurements (including the
location of the central sites $s_c$ as defined in Eq.~\eqref{eq:centsite}). This is due to the fact
that the $E_n$ are scattered above and below the exact value, resulting in an average that is
more accurate. Although isolated sites more accurate than the average value exist, these do not
occur at specific locations independent of~$m$, which is why this behavior has to be
attributed to numerical fluctuations rather than to some systematic reason.
Therefore, Fig.~\ref{fig:ising_averaging} strongly suggests that averaging over all sites is the 
best way to extract local observables from a 1D-TTN.
We proceed to analyze the averaging issue for 
correlation functions. Since it is efficient to calculate correlators $C^\alpha_{n,n+r}$
for every distance~$r$ and pair position~$n$,
we can easily address the question whether or not it is profitable to calculate the average
$\sum_{n=1}^N C^\alpha_{n,n+r}/N$ for a given pairwise distance~$r$.
Fig.~\ref{fig:ising_averaging} shows the error $\Delta
C^y_{n,n+r}$ of the spin correlations in y-direction as a function of $n$ and~$r$. Immediately
noticeable is a triangular pattern, indicating that with increasing~$r$ the accuracy abruptly
degrades each time a major branch in the tree is trespassed. 
The zoomed panels show that for this reason it 
is advantageous (especially for short distances), to measure the correlator within a branch wide
enough to span the given distance, i.e.~within the dark blue areas of figure Fig.~\ref{fig:ising_averaging}.
Averaging, in this case, is mainly detrimental and results in 
significantly bigger errors, up to two orders of magnitude.

\section{Bilinear-biquadratic spin-1 chain}

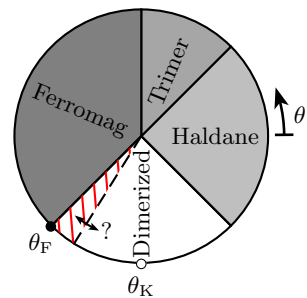
\begin{figure}[b]
	\newlength{\diagrad} \setlength{\diagrad}{1.7cm}
\begin{pspicture}(-4,-1.2\diagrad)(4,\diagrad)
    \pswedge[fillstyle=solid, fillcolor=lightgray]{\diagrad}{-45}{45}
    \pswedge[fillstyle=solid, fillcolor=mediumgray]{\diagrad}{45}{90}
    \pswedge[fillstyle=solid, fillcolor=gray]{\diagrad}{90}{225}
    \pswedge{\diagrad}{-135}{-45}
	\pswedge[linestyle=none,fillstyle=vlines, hatchangle=5, hatchcolor=red]{\diagrad}{-135}{-122}
	\psline[linestyle=dashed](\diagrad;-122)(0;0)    
    
	\rput[l]{0}(0.4;0){\small Haldane}
	\rput[l]{67.5}(0.45;67.5){\small Trimer}
	\rput[l]{-22.5}(1.55;157.5){\small Ferromag}
	\rput[l]{90}(1.62;-90){\small Dimerized}
	
	\psarc[arrowsize=0.15 1.4, arrowinset=0.25, linewidth=1pt]{|->}{1.9}{0}{20}
	\rput[l](2.05;8){$\theta$}
	\psdot[dotsize=4pt](\diagrad;-135)
	\rput(1.95;-133){$\theta_\mathrm{F}$}
	\psdot[dotsize=4pt, dotstyle=Bo](\diagrad;-90)
	\rput(2.0;-90){$\theta_\mathrm{K}$}
	\psarc[linestyle=dashed,arrowsize=0.1,arrowinset=0.2]{<->}{0.81\diagrad}{-130}{-114}
	\rput(0.79\diagrad;-110){?}
\end{pspicture}
	\caption{\label{fig:blbq_phasediag}
	Phase diagram of the bilinear-biquadratic model of Eq.~\eqref{eq:ham_blbq},
	with four firmly established phases: ferromagnetic, trimerized~\cite{Laeuchli2006},
	Haldane antiferromagnet~\cite{Haldane1983, Haldane1983a} and dimerized~\cite{Barber1989,Xian1993}.
	According to the most recent studies and the data obtained here with adaptive TTN,
	no intermediate nematic phase (originally hypothesized location indicated by red  
	hatched region~\cite{Chubukov1990,Chubukov1991}) 
	is emerging around $\theta_\mathrm{F}$ (black dot).
	The white dot identifies the Kl\"umper point~\cite{Barber1989,Kluemper}
	whose Bethe ansatz solution in Eq.~\eqref{eq:energ_biq} we use to benchmark the 
	numerical precision of our algorithm.}
\end{figure}

After having benchmarked the adaptive TTN algorithm on the spin-$1/2$ quantum Ising model,
in this section we turn to a numerically very challenging transition in $SU(2)$ invariant spin-$1$ chains.
We consider indeed the bilinear-biquadratic Hamiltonian ($H \equiv \sum_{n=1}^{N-1} h_{n,n+1}$):
\begin{equation}
	H = \sum_{n=1}^{N-1} \cos\theta \left( \vec{S}_n \cdot \vec{S}_{n+1} \right) + 
		\sin\theta \left( \vec{S}_n \cdot \vec{S}_{n+1} \right)^2 \; ,
	\label{eq:ham_blbq}
\end{equation}
whose phase diagram~\cite{Papanicolaou1988, Fath1995} is provided in Fig.~\ref{fig:blbq_phasediag}
as a function of the parameterizing angle $\theta$.
While most phase boundaries have been set on firm grounds relatively early%
~\cite{Haldane1983, Haldane1983a,Barber1989,Xian1993,Laeuchli2006},
the region between the translationally broken dimer and the rotationally broken ferromagnet
have been subject of a long and controversial debate%
~\cite{Chubukov1990, Chubukov1991,Fath1995, Kawashima2002, Ivanov2003,
Buchta2005, Rizzi2005, Porras2006, Laeuchli2006, Orus2011}. 
An intermediate nematic phase, restoring the translational invariance while mildly breaking the rotational one
by a quadrupolar moment, has been conjectured~\cite{Chubukov1990, Chubukov1991}
in the range $-3\pi/4 \equiv \theta_\mathrm{F} \leq \theta \leq \theta_\mathrm{C} \simeq -0.67\pi$.
%%%%%%%%%%%%%%%%%%%
%%%%%%%%%%%%%%%%%%%
The competing order parameters would then be:
the dimerization $D$~\cite{Xian1993}
\begin{equation}
  D_n = |\langle h_{n-1,n} - h_{n,n+1} \rangle| \; ,
  \label{eq:dim_param}
\end{equation}
which can be non-zero in 1D finite chains;
and a quadrupole moment $Q_{\Omega}$ (with $\vec{e}_{\Omega}$ a unit vector 
pointing to the solid angle $\Omega$)
\begin{equation}
Q_{n,\Omega}=( \vec{S}_n \cdot \vec{e}_{\Omega} )^2 - 2/3 \; ,
  \label{eq:quad_param}
\end{equation}
of which only quasi-long range order in correlations can be 
measured in 1D finite chains (see Eq.~\eqref{eq:quadcorr}).
Increasingly powerful numerical techniques%
~\cite{Fath1995, Kawashima2002, Ivanov2003, Buchta2005, Rizzi2005, Porras2006, Laeuchli2006, Orus2011}
and improved theoretical analyses~\cite{Hu2014,Moessner,Grover}
have shrunk more and more the nematic window, by accumulating evidences of
an exponentially vanishing $D$ accompanied by an exponentially growing 
quadrupole correlation length $\xi_\mathrm{Q}$.
The latter accounts for the extreme difficulty in ruling out a quasi-long range nematic order
only based on numerical data of finite-size chains.
A recent description of the Berry phases of quantum fluctuations~\cite{Hu2014} 
predicted the following scalings with $\Delta \theta = \theta - \theta_\mathrm{F}$:
\begin{subequations}
\label{eq:scal_behav}
\begin{eqnarray}
  D \approx && \exp \left( - \pi^2 / (8 \Delta \theta) \right) \; ,
  \label{eq:scal_dimer} \\
  \xi_\mathrm{Q} \approx && \exp \left( \pi \sqrt{2/\Delta \theta} \right) \; . 
  \label{eq:scal_corrlen}
\end{eqnarray}
\end{subequations}
The data we obtain from our adaptive TTN procedure actually nicely agree with this prediction, as we
argue below, and thus corroborate the validity of the presented method also in non-trivial cases.

\begin{figure}[b]
	\input{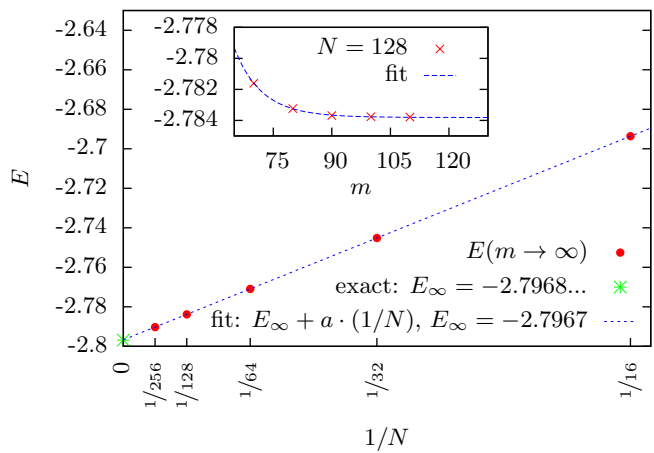}
	\\[5mm]
	\caption{\label{fig:blbq_OBC_ener}Extrapolation to the thermodynamic limit of the ground state 
	energy at the purely biquadratic point $\theta=-\pi/2$. The inset shows the extrapolation in the
	bond dimension for $N=128$, using a fit similar to the one indicated in
	Fig.~\ref{fig:ising_energyerr}.}
\end{figure}

First, we assess the numerical precision of the adaptive TTN algorithm 
by analysing its performances in the purely 
biquadratic Kl\"umper point~\cite{Kluemper} 
at $\theta = \theta_\mathrm{K} = -\pi/2$ (empty circle in Fig.~\ref{fig:blbq_phasediag}).
The ground state energy in the thermodynamic limit can be  
analytically obtained by Bethe Ansatz 
(after a mapping to a spin-1/2 XXZ chain) as~\cite{Barber1989,Soerensen1990}
\begin{align}
  E/N =& -1 - 2 \, \sqrt{5} \; \left( \frac{1}{4} + \sum_{k=1}^\infty \;\,
  	\frac{1}{ 1 + \left[ \left( 3+\sqrt{5}\right)/2 \right]^{2k} } \right) \nonumber \\
  	=& -2.79686343... \;  ,
  \label{eq:energ_biq}
\end{align}
and will serve as a reference for our finite-bond and finite-size extrapolation procedure.
This consists in taking first the limit $m \rightarrow \infty$ for every chain length $N$
(according to the empirical formula indicated in Fig.~\ref{fig:ising_energyerr}), 
as illustrated in the 
inset of Fig.~\ref{fig:blbq_OBC_ener}, and then considering the 
thermodynamic limit $N \rightarrow \infty$ by a linear fit in $1/N$
to eliminate edge effects, as shown in the main plot of Fig.~\ref{fig:blbq_OBC_ener}.
The thus estimated energy $E_\mathrm{TTN} = -2.79670(1)$ has a precision of $10^{-4}$.
%%%%%%%%%%%%%%%%%%%%%%%%%%%%%%%%%%%%%%%%%%%%%%%%%%
%%%%%%%%%%%%%%%%%%%%%%%%%%%%%%%%%%%%%%%%%%%%%%%%%%

\begin{figure}
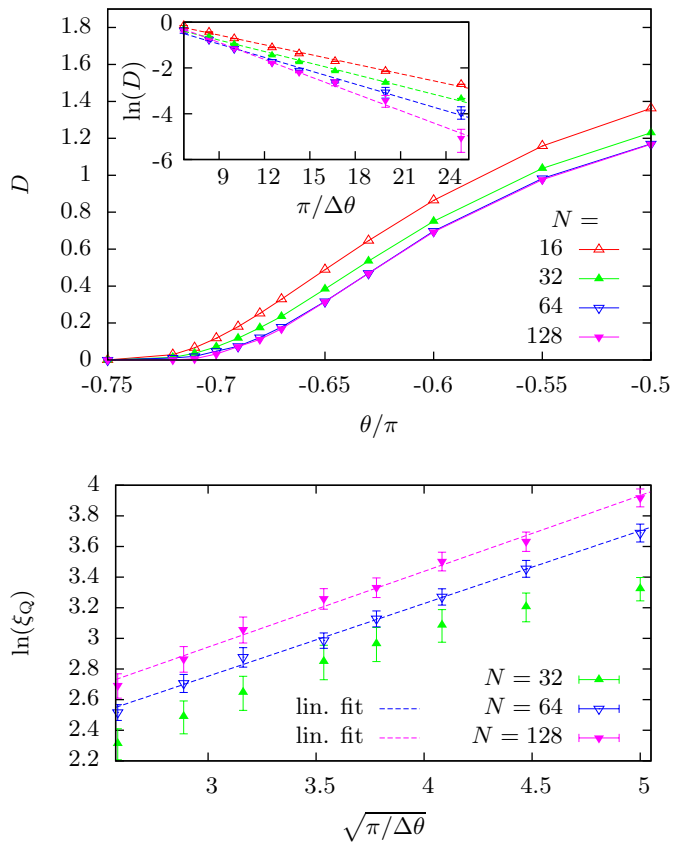

	\hspace{-3mm}\input{gnuplottex-fig8}
	\\[5mm]
	\hspace{-3mm}\input{gnuplottex-fig9}
	\\[5mm]
	\caption{\label{fig:blbq_OBC_dimerAndcorrlen}
	(Top) Dimerization parameter $D$ for various system sizes.
	Each data point is the result of an extrapolation $m~\rightarrow~\infty$ in the bond dimension.
	The dashed lines shown in the inset are linear fits.
	(Bottom) Quadrupolar correlation length $\xi_{\mathrm{Q}}$, obtained from fitting the correlation
	function Eq.~\eqref{eq:quadcorr}. Each $C^{\mathrm{Q}}(r)$ has been extrapolated to $m\rightarrow \infty$
	(for every $r$) before the correlation length was extracted. Reported error bars are fit errors. The large
	error bars for $N=32$ are caused by the fact that the chain is too small to contain meaningful
	information on the long-range property $\xi_{\mathrm{Q}}$.
	}
\end{figure}

We then proceed to evaluate, for various values of $\theta \in [\theta_\mathrm{F}, \theta_\mathrm{K}]$,
the dimerization $D$ of Eq.~\eqref{eq:dim_param}.
In order to avoid issues related to resonant superpositions of different dimerized configurations,
we resort to open boundary conditions. We avoid end effects of the open chain by measuring the
dimerization only at sites that are sufficiently far away from the edges.
The data presented in the upper plot of Fig.~\ref{fig:blbq_OBC_dimerAndcorrlen}
nicely agree with the prediction of Eq.~\eqref{eq:scal_dimer}, 
displaying a clear exponential behavior in~$1 / \Delta\theta$. 

Finally, we proceed to compute the correlation of quadrupole moments at a long distance $r$
averaged over the solid angle: 
\begin{eqnarray}   \label{eq:quadcorr} 
  C^\mathrm{Q}(r) & = & \int \frac{\mathrm{d}\Omega}{4\pi} \langle Q_{n, \Omega} \, Q_{n+r, \Omega} \rangle 
  - \langle Q_{n, \Omega} \rangle \langle Q_{n+r, \Omega} \rangle \\
  & = & \frac{2}{15} \sum_\alpha \left\langle \left( S^\alpha_{n} \right)^2 \left( S^\alpha_{n+r} \right)^2 \right\rangle
  - \left\langle \left( S^\alpha_{n} \right)^2 \right\rangle \left\langle \left( S^\alpha_{n+r} \right)^2 \right\rangle
  \nonumber \\
  & + & \frac{1}{15} \sum_{\alpha < \beta} \left\langle T^{\alpha \beta}_{n} T^{\alpha \beta}_{n+r} \right\rangle
  - \left\langle  T^{\alpha \beta}_{n}  \right\rangle \left\langle  T^{\alpha \beta}_{n+r} \right\rangle \; ,
  \nonumber
\end{eqnarray}
with the anticommutator $T^{\alpha\beta}_n \equiv \left\{S^\alpha_n, S^\beta_n \right\}$.
The correlation length $\xi_\mathrm{Q}$ is then extracted by fitting the $m \rightarrow \infty$ 
extrapolations with $C^\mathrm{Q}(r) = a \, r^{-\eta} \, \exp(-r/\xi_{\mathrm{Q}})$, 
where $a$ is a prefactor and $\eta$ is an exponent. 
The results are shown in the lower panel of Fig.~\ref{fig:blbq_OBC_dimerAndcorrlen}
and provide quite clear evidence of an 
exponential growth of $\xi_\mathrm{Q}$ with $1/\sqrt{\Delta\theta}$.
In summary, these results support Eqs.~\eqref{eq:scal_behav} and therefore indicate 
the non-existence of the nematic phase.
%%%%%%%%%%%%%%%%%%%%%%%%%%%%%%%%%%%%%%%%%%%%%%%%%%
%%%%%%%%%%%%%%%%%%%%%%%%%%%%%%%%%%%%%%%%%%%%%%%%%%

\section{Conclusions}

In this manuscript we introduced, motivated, and discussed an algorithm for simulating ground states
of quantum many-body Hamiltonians on a lattice, based on a gauge-adaptive tree tensor network
ansatz.
We stressed how the manipulation at runtime of the tensor network gauge, combined with the loop-free
topology of the tree network, plays an instrumental role in enhancing 
the performance of the algorithm,
first of all, by reducing a generalized eigenvalue problem into a simple eigenvalue problem.
We characterized the computational scalings with bond dimension $m$ and argued how the
algorithm speed-up allows for pushing to higher numerical precision.
We tested the algorithm on 1D quantum models, with various boundary conditions:
first we benchmarked it against the exactly solvable Ising model in a transverse field,
extracting very precise critical exponents with PBC's;
then we explored the bilinear-biquadratic \mbox{spin-1} model in a region known to be  
numerically challenging, being able to confirm an intricate exponential explosion of  
(nematic) correlation lengths towards the transition point. 
Moreover, we investigated how well the TTN ansatz can restore
the broken translational invariance in 1D, and verified that while for local quantities averaging
over translations is a winning strategy, for correlation functions evaluating over highly-entangled
clusters is a more suitable approach.

The redirectionable isometric gauge presented here allows also for a convenient, 
natural treatment of preserved global symmetries,
both abelian and non-abelian, which result in a block-diagonal structure of the tensor 
entries with clear, unique selection rules 
and structure constants~\cite{SymTN1, SymTN2, Psitesi}
(in this framework, a simultaneous double-tensor optimization scheme would help circumventing the
related intermediate charge-targeting issues\cite{Whiteonesite}).
Similarly, lattice gauge symmetries, which in contrast to global symmetries
act on local compact supports of the 1D chain of sites, 
can be rigorously and efficiently cast in tensor network 
language~\cite{EnriqueLGT,Buyens2013,Silvi2014,Tagliacozzo2014}, 
and in particular in a TTN ansatz.

An additional research direction is offered by the recently introduced time dependent variational
principle for tensor network states~\cite{JuthoTDVP}, whose purpose is to describe efficiently
out-of-equilibrium dynamics: indeed it can be elegantly and successfully tailored to TTN, free of
tensor gauge complications thanks to the loop-free geometry of the tree network.
 
Finally, we stress again that the flexibility of the gauge-adaptive TTN ansatz presented here, and
especially the fact that its scaling is independent from the specific (open or periodic) boundary
conditions, makes it an extremely promising tool to attack and solve open complex many-body
problems spanning from condensed matter to quantum information. Moreover, the presented approach
can be trivially extended to include lattice dimensions higher than one with the same polynomial
scaling of the algorithm’s computational cost.

ACKNOWLEDGEMENTS - We thankfully acknowledge J.I. Cirac for inspiring this work, for feedback and
discussions.
Authors acknowledge financial support from EU through SIQS, the German Research Foundation (DFG) via
the SFB/TRR21, Italian MIUR via PRIN Project 2010LLKJBX, and computational resources provided by the
bwUniCluster and bwGRiD projects in Ulm and by the MOGON cluster of the ZDV Data Center in Mainz.

\bibliography{references}% Produces the bibliography via BibTeX.

\end{document}